\documentclass[useAMS]{mn2e}
\usepackage{graphicx}
\def\gsim{\mathrel{\raise.5ex\hbox{$>$}\mkern-14mu
             \lower0.6ex\hbox{$\sim$}}}
\def\lsim{\mathrel{\raise.3ex\hbox{$<$}\mkern-14mu
             \lower0.6ex\hbox{$\sim$}}}

\title[RT]{The Magnetic Rayleigh-Taylor Instability in Astrophysical Disks}
\author[Papadopoulos, Contopoulos \& Kazanas]{I. Contopoulos$^{1,2}$\thanks{E-mail:
icontop@academyofathens.gr}, D. Kazanas$^{3}$
and D. B. Papadopoulos$^{4}$\\
$^{1}$Research Center for Astronomy and Applied Mathematics,
Academy of Athens, Athens 11527, Greece\\
$^{2}$National Research Nuclear University, Moscow 115409, Russia\\
$^{3}$NASA Goddard Space Flight Center, Laboratory for High-Energy
Astrophysics, Code 663, Greenbelt, MD 20771, USA\\
$^{4}$Department of Physics, Aristotle University of Thessaloniki,
Thessaloniki 54124, Greece}
\begin{document}

\date{Accepted ... Received ...; in original form ...}

\pagerange{\pageref{firstpage}--\pageref{lastpage}} \pubyear{2015}

\maketitle

\label{firstpage}

\begin{abstract}
This is our first study of the magnetic Rayleigh-Taylor instability at
the inner edge of an astrophysical disk around a central back hole.
We derive the equations governing small-amplitude oscillations in
general relativistic ideal magnetodydrodynamics and obtain a
criterion for the onset of the instability. We suggest that static disk configurations
where magnetic field is held by the
disk material are unstable around a Schwarzschild black hole.
On the other hand, we find that such configurations are stabilized by the spacetime
rotation around a Kerr black hole. We obtain a crude estimate of the
maximum amount of poloidal magnetic flux that can be
accumulated around the center, and suggest that it is
proportional to the black hole spin. Finally, we discuss the astrophysical
implications of our result for the theoretical and observational estimations
of the black hole jet power.
\end{abstract}


\section{Introduction}

Astrophysical magnetic fields are believed to play a fundamental
role in powering astrophysical energetic sources such as active
galactic nuclei, X-ray binaries, and gamma-ray bursts.
Extensive research over the last 4 decades has most
convincingly shown that magnetic fields contribute to the
extraction of rotational energy from astrophysical accretion disks
and compact objects (neutron stars, black holes), and to the
launching, collimation and acceleration of astrophysical jets. The
electrodynamically extracted power is proportional to the square
of the total amount of open magnetic flux that threads the central
spinning compact object. In the case of a spinning neutron star,
the magnetic field originates in the stellar interior and is held
in place by the highly conducting neutron star matter. In the case
of a spinning black hole, however, the magnetic field is held in
place by the surrounding disk of matter, and if the disk is
removed, the magnetic field escapes the system at light crossing
times.

The origin of the large scale astrophysical magnetic field held by
the accretion disk around a spinning black hole is not clear. One
school of thought suggests that the field is brought in from large
scales by the accretion flow, and several numerical simulations
are set up with a `reservoir' of large scale poloidal magnetic
flux at large distances (e.g. Tchekhovskoy, Narayan \& McKinney
2011). The main problem with this scenario is that astrophysical
accretion disks are viscous, thus also diffusive, and therefore
they can hardly advect any magnetic flux over so many orders of
magnitude in radius (e.g. Lubow et al.~1994). To our
understanding, the problem of how magnetic flux is brought in from
large distances is still open (e.g. Lovelace et al.~2009; Kylafis
et al.~2011). Another more promising astrophysically plausible
scenario proposes that the magnetic field is generated around the
inner edge of the accretion disk. This is the Cosmic Battery
according to which, one polarity is advected inward and inundates
the black hole horizon, whereas the return polarity diffuses
outward through the surrounding disk (Contopoulos \& Kazanas~1998;
Contopoulos, Nathanail \& Katsanikas~2015).

Whatever the origin of the magnetic field turns out to be, the
common understanding is that the
collected field is held in place by the `weight'
of the disk that keeps the magnetic field from escaping.
According to this understanding,
the growth of the field cannot continue beyond a so
called equipartition limit $B_{\rm eq}$ where the magnetic field
energy density either balances the accretion disk ram pressure,
namely
\begin{equation}
\frac{B_{\rm eq}^2}{8\pi} \sim \dot{M}_{\rm disk}\frac{v_{\rm
K}}{4\pi r^2}\ ,
\label{1}
\end{equation}
or balances the full weight of the inner disk, namely
\begin{equation}
\frac{B_{\rm eq}^2}{8\pi} \sim \frac{GM M_{\rm
disk}}{4\pi r^4}
\label{2}
\end{equation}
(eq.~\ref{2} follows from eq.~\ref{1} for thick disks only).
Here, $M$ is the mass of the central black hole. When the
magnetic field (or equivalently the total accumulated magnetic flux) reaches
a value on the order of the
above limits, accretion will be disrupted. Such configuration is termed Magnetically
Arrested Disk (MAD; Igumenshchev 2008). Recent state-of-the-art numerical
simulations have shown the MAD process in action. In axisymmetry
(2D), when the accumulated magnetic field reaches the
above maximum value, accretion stops. In realistic 3D accretion
though, magnetic flux can escape the system in the
azimuthal-$\phi$ direction as shown very clearly in the numerical
simulations of e.g. Tchekhovskoy et al.~(2011). The
breaking of the axisymmetry by the azimuthal `bunching up' of the field lines is precisely the
{\em magnetic Rayleigh-Taylor instability}. And here rises the obvious
question: how stable are MAD configurations against this instability?

In classical fluid motion, the Rayleigh-Taylor instability has been investigated
by several authors in both hydrodynamics and magnetohydrodynamics
(Chandrasekhar~1961; Kruskal and Schwarzschild~1954;
an interesting presentation can be found in Boyd and Sanderson~1969).
The aim of the present work is to determine more precisely the main
parameters that characterize the onset of this important
instability around astrophysical black holes.
Numerical simulations yield the amount of magnetic flux that is
effectively held around the central spinning black hole which, as
we said, is a fundamental parameter that determines the efficiency
of energy production is energetic astrophysical sources. We would
like to be able to obtain the same result from first principles.
This will allow us to determine whether an astrophysical black
hole is active (implying that it generates jets that extract
energy from its rotation) or inactive. Another very important future
application of the present work would be to explain the various
stages of a flaring X-ray binary where too, as shown in Kylafis et
al.~(2012) the main parameter that characterizes the evolution is
the generation and destruction of the large scale magnetic flux
accumulated around the black hole horizon.

The goal of this paper is to obtain the magnetic Rayleigh-Taylor
stability criterion for an astrophysical disk with a central black hole.
We were able to achieve our goal only in the simplified case
of two static distributions of
ideal magnetized plasma in contact with each other
in the equatorial plane of the central black hole.
We perturbed the contact interface in the
radial and azimuthal direction and considered a particular form of
velocity perturbations that allowed us to obtain a
criterion for the stability of the interface. In the next section we develop our
general relativistic formalism, and in \S~3 we apply it to obtain the general
stability criterion on the equatorial plane. In the next two sections we apply our results
around non-rotating and rotating black hole respectively, and
in the final section, we discuss the astrophysical implications
of our work.

\section{General Relativistic MHD in 3+1 formalism}

We will follow the 3+1 (space+time) formalism of general
relativistic magnetohydrodynamics (GRMHD) of Thorne \&
Macdonald~(1982). We introduce spatial magnetic and electric
fields (${\bf B}$ and ${\bf E}$ respectively) as measured by
fiducial observers with 4-velocity $U^{\mu}$. In that formalism,
Maxwell's equations $F_{;\beta}^{\alpha\beta}=4\pi J^{\alpha}$,
$F_{[\alpha\beta;\gamma]}=0$, and $J_{;\alpha}^{\alpha}=0$ yield
\begin{eqnarray}
&&\tilde{\nabla}\cdot \tilde{E} = 4\pi\rho_e\nonumber\\
&&\tilde{\nabla}\cdot \tilde{B} = 0\nonumber\\
&&D_{\tau}\tilde{E}+\frac{2}{3}\theta
\tilde{E}-\tilde{\sigma}\cdot
\tilde{E} = \frac{1}{\alpha}\tilde{\nabla}\times (\alpha \tilde{B})-4\pi \tilde{J}\nonumber\\
&&D_{\tau}\tilde{B}+\frac{2}{3}\theta
\tilde{B}-\tilde{\sigma}\cdot \tilde{B} =
-\frac{1}{\alpha}\tilde{\nabla}\times (\alpha \tilde{E})\label{k9}
\label{fullMHD1}
\end{eqnarray}
with
\begin{eqnarray}\label{k10}
&&D_{\tau}\rho_e+\theta \rho_e+\frac{1}{\alpha}\tilde{\nabla}\cdot
(\alpha \tilde{J})=0\ .
\end{eqnarray}
Here, $D_{\tau} M^{\beta}\equiv M^{\beta}~_{;\mu}
U^{\mu}-U^{\beta}a_{\mu} M^{\mu}$ is the Fermi derivative,
$\theta$ and $\tilde{\sigma}$ are the expansion and shear of the
spacetime metric respectively. The evolution of the magnetized
fluid is characterized by the divergence of the total
stress-energy tensor $T^{\mu\nu}\equiv T_{\rm
matter}^{\mu\nu}+T_{\rm EM}^{\mu\nu}$, namely
\begin{eqnarray}
&&T^{\mu\nu}~_{;\nu}=0\ ,
\end{eqnarray}
which yields
\begin{eqnarray}
&&D_{\tau}\varepsilon+\theta\varepsilon+\frac{1}
{\alpha^2}\tilde{\nabla}\cdot(\alpha^2 \tilde{S})
+W^{jk}(\sigma_{jk}+\frac{1}{3}\theta\gamma_{jk})=-\tilde{J}\cdot
\tilde{E}\nonumber\\
&&D_{\tau}\tilde{S}+\frac{4}{3}\theta
\tilde{S}+\tilde{\sigma}\cdot\tilde{S}+\varepsilon
\tilde{a}+\frac{1}{\alpha}\tilde{\nabla}\cdot(\alpha
\tilde{W})=\nonumber\\
&&\rho_e\tilde{E}+\tilde{J}\times\tilde{B}\ .\label{k12}
\end{eqnarray}
Here,
\begin{eqnarray}\label{k13}
&&\varepsilon\equiv T_{\rm matter}^{\mu\nu}U_{\mu}U_{\nu}\nonumber\\
&&S^{\alpha}\equiv \gamma^{\alpha}~_{\mu} T_{\rm matter}^{\mu\nu}
U_{\nu}\nonumber\\
&&W^{\alpha\beta}\equiv \gamma^{\alpha}~_{\mu} T_{\rm matter}{\mu\nu} \gamma^{\beta}~_{\nu}\nonumber\\
&&\theta\equiv U^{\mu}~_{;\mu},~~ a^{\mu}\equiv
U^{\mu}~_{;\nu}U^{\nu}
, ~~\nonumber\\
&&\sigma_{ab}\equiv \frac{1}{2}\gamma^{\mu}~_{a}\gamma^{\nu}~_{b}(U_{\mu;\nu}+U_{\nu;\mu})-\frac{1}{3}\theta \gamma_{ab}\nonumber\\
&&\tilde{L}\cdot \tilde{M}=\gamma^{ij} L_i
M_j,~~(\tilde{L}\times\tilde{M})^j=\epsilon^{ijk} L_j M_{\rm K}\ ,
\end{eqnarray}
and, $\gamma^{\alpha\beta}=g^{\alpha\beta}+U^\alpha U^\beta$ is
the projection tensor, and $\alpha$ is the lapse function. Latin
indices take values $1,2,3$ and Greek ones $0,1,2,3$. Vectors and
tensors with tildae are purely spatial. For an ideal fluid with
density $\rho$, 3-velocity $\tilde{v}$, and pressure $p$ we have
\begin{eqnarray}\label{k14}
&&\Gamma=(1-\tilde{v}^2)^{-1/2},~~\varepsilon= \Gamma^2(\rho+p\tilde{v}^2)\nonumber\\
&&\tilde{S}=(\rho+p)\Gamma^2
\tilde{v},~~\tilde{W}=(\rho+p)\Gamma^2\tilde{v}\otimes\tilde{v}+p\tilde{\gamma}\
.
\end{eqnarray}
We will also assume an equation of state $p=p(\rho)$ from which we
deduce the `speed of sound'
\begin{eqnarray}
&&c_s\equiv \left(\frac{dp}{d\rho}\right)^{1/2}\ .
\end{eqnarray}
Finally, we will also assume ideal MHD conditions, namely
\begin{eqnarray}\label{l1x}
&&\tilde{E}=-\tilde{v}\times \tilde{B}
\end{eqnarray}
In order to investigate the Rayleigh-Taylor instability in an
astrophysical context, we will now consider the special case of a
Kerr space-time.

\subsection{Kerr spacetime}

In Boyer-Lindquist (BL) coordinates the Kerr metric reads
\begin{eqnarray}\label{k1x}
ds^2 & = & g_{tt} dt^2+2g_{t\phi}dt d\phi +g_{rr}
dr^2+g_{\theta\theta} d\theta^2+g_{\phi\phi} d\phi^2\nonumber\\
& = & -(1-\frac{2M r}{\Sigma})dt^2-\frac{4 M a
r\sin^2{\theta}}{\Sigma}dt d\phi\nonumber\\
&&+\frac{\Sigma}{\Delta} dr^2+\Sigma
d\theta^2+\frac{A}{\Sigma}\sin^2{\theta}d\phi^2\end{eqnarray}
where $M$ is the mass of the black hole, $a$ is the angular
momentum per unit mass $(0\leq a \leq M)$, and
\begin{eqnarray}\label{k2x}
&&\Delta \equiv r^2-2M r+a^2\nonumber\\
&&\Sigma \equiv r^2+a^2\cos^2{\theta}\nonumber\\
&&A \equiv(r^2+a^2)^2-a^2\Delta\sin^2{\theta}\end{eqnarray}
(Cowling~1941). Notice that we work in geometrical units in which
$c=G=1$.

For our further study  we need the components of the 4-velocity of
fiducial observers, now identified as ZAMOs (Zero Angular Momentum
Observers), namely
\begin{equation}\label{k3x}
U^{\mu}=(\frac{1}{\alpha},0,0,\frac{\omega}{\alpha})\ ,\ \
U_{\mu}=(-\alpha,0,0,0)
\end{equation}
where
\begin{equation}\label{k5x}
\alpha=\sqrt{\frac{\Delta\Sigma}{A}},~~\omega=\frac{2M
ar}{A}\end{equation}
In the Kerr spacetime (\ref{k1x}) with 4-velocity $U^{\mu}$ given
by eq.~(\ref{k3x}), the expansion $\theta$ vanishes, the shear $\tilde{\sigma}$
has two non-zero components e.g. the $\sigma^{13}$ and $\sigma^{23}$ but
$\sigma_{\alpha\beta}\gamma^{\alpha \beta}=0$; (see Thorne \&
Macdonald~1982, paper~I, eq.~(2.5)). The acceleration $a^{\mu}$ is
given by
\begin{eqnarray}\label{k6ax}
a^{\mu}&=&(0,\frac{-M a^2\cos^2{\theta}[(r^2+a^2)^2-4Mr^3]}
{\Sigma^2 A}\nonumber\\
&&+\frac{Mr^2[(a^2+r^2)^2-4Mra^2]}{\Sigma^2 A}\nonumber\\
&&,\frac{M r a^2(r^2+a^2)\sin{2\theta}}{\Sigma^2 A},0)\ .
\end{eqnarray}
$\gamma_{ij}$ is the spatial metric on the space-like hypersurface
$x^0\equiv t=$~const., with normal vector $n_{\alpha}$
\begin{equation}\label{k8x}
n_{\alpha}=(-\alpha,0,0,0),~~n^{\alpha}=\frac{1}{\alpha}(1,-\beta^1,-\beta^2,-\beta^3)\end{equation}
where $\beta^i=\gamma^{ij}g_{0j}$.

\subsection{The perturbed MHD equations}

We consider only small perturbations of physical quantities as
\begin{eqnarray}
&&\rho(t,\tilde{r})=\rho_0(\tilde{r})+\delta\rho(t,\tilde{r})\nonumber\\
&&\rho_e(t,\tilde{r})=\rho_{e0}(\tilde{r})+\delta\rho_e(t,\tilde{r})\nonumber\\
&&v^{i}(t,\tilde{r})=v^{i}_0(\tilde{r})+\delta v^{i}(t,\tilde{r})\nonumber\\
&&B^{\mu}(t,\tilde{r})=B_0^{\mu}(\tilde{r})+\delta B^{\mu}(t,\tilde{r})\nonumber\\
&&E^{\mu}(t,\tilde{r})=E_0^{\mu}(\tilde{r})+\delta E^{\mu}(t,\tilde{r})\nonumber\\
&&J^{\mu}(t,\tilde{r})=J_0^{\mu}(\tilde{r})+\delta
J^{\mu}(t,\tilde{r})\label{per1}
\end{eqnarray}
and keep only linear terms of the perturbations. In this case
\begin{eqnarray}\label{v1}
v^2&=&\gamma_{ij}v^iv^j=\gamma_{ij}(v_0^i+\delta v^i)(v_0^j+\delta v^j)\nonumber\\
&=&\gamma_{ij}v_0^iv_0^j+2\gamma_{ij}v_0^i\delta v^j\nonumber\\
&=&\gamma_{rr}(v_0^r)^2+\gamma_{\phi\phi}(v_0^{\phi})^2\nonumber\\
& & +2\gamma_{rr}v_0^r\delta
v^r+2\gamma_{\phi\phi}v_0^{\phi}\delta v^{\phi}
\end{eqnarray}
and
\begin{eqnarray}\label{v2}
\Gamma^2&=&\{1-[\frac{\Sigma}{\Delta}(v_0^r)^2+\frac{A}{\Sigma}\sin^2{\theta}(v_0^{\phi})^2\nonumber\\
&&+2\frac{\Sigma}{\Delta}v_0^r\delta
v^r+2\frac{A\sin^2{\theta}}{\Sigma} v_0^{\phi}\delta
v^{\phi}]\}^{-1}
\end{eqnarray}
In the Cowling approximation of a fixed Kerr spacetime,
\begin{eqnarray}\label{mhd7}
\delta \Gamma^2 & = & \frac{v_0^k\delta v_{\rm K}}{(1-v^2)^2}+\frac{v_{0k}\delta v^k}{(1-v^2)^2}=2\frac{\gamma_{kl}v_0^k\delta v^l}{(1-v^2)^2}\nonumber\\
\delta\varepsilon & = & 2(\rho+v^2 p)\frac{v_0^k \delta
v_{\rm K}}{(1-v^2)^2}\nonumber\\
&&+\frac{1}{1-v^2}[\delta \rho+v_0^2\delta p+2p v_0^k\delta
v_{\rm K}]\nonumber\\
\delta S^i & = & \frac{v_0^i}{1-v^2}(\delta \rho+\delta p)
+2\frac{v_0^k\delta v_{\rm K}}{(1-v^2)^2}(\rho+p)v_0^i\nonumber\\
&&+\frac{\delta v^i}{1-v^2}(\rho+p)\nonumber\\
\delta W^{ij} & = & \frac{v_0^i v_0^j}{1-v^2}(\delta \rho+\delta p)
+2\frac{v_0^k\delta v_{\rm K}}{(1-v^2)^2}(\rho+p)v_0^iv_0^j\nonumber\\
&&+\frac{(\rho+p)}{1-v^2}(v_0^j\delta v^i+v_0^i\delta v^j)+\gamma^{ij}\delta p\nonumber\\
\label{mhd8}\end{eqnarray}
The first order perturbed MHD equations now become
\begin{equation}\label{mhd1a}
\tilde{\nabla}\cdot \delta\tilde{E}=4\pi\delta\rho_e,
\end{equation}
\begin{equation}\label{mhd1b}
\tilde{\nabla}\cdot \delta\tilde{B}=0,
\end{equation}
\begin{equation}\label{mhd1c}
D_{\tau}\delta\tilde{E}=\tilde{\nabla}\times \delta\tilde{B}+\tilde{a}\times \delta \tilde{B}+\tilde{\sigma}\cdot \delta\tilde{E}-4\pi \delta\tilde{J},
\end{equation}
\begin{equation}\label{mhd1d}
D_{\tau}\delta \tilde{B}=-\tilde{\nabla}\times \delta\tilde{E}-\tilde{a}\times \delta\tilde{E}+\tilde{\sigma}\cdot\delta\tilde{B},
\end{equation}
\begin{equation}\label{mhd1e}
D_{\tau}\delta\rho_e+\delta\tilde{J}\cdot \tilde{a}+\tilde{\nabla}\cdot \delta\tilde{J}=0,
\end{equation}
\begin{equation}
D_{\tau}\delta\rho +2\delta\tilde{S}\cdot \tilde{a}+\tilde{\nabla}\cdot \delta\tilde{S}+\tilde{\sigma}\cdot\delta \tilde{W}
=-\delta\tilde{J}\cdot\tilde{E}_0-\tilde{J}_0\cdot\delta\tilde{E}\ ,\\
\label{mhd1f}
\end{equation}
\begin{equation}
D_{\tau}\delta\tilde{S}+\tilde{a}\delta\rho +\delta\tilde{W}\cdot \tilde{a}+\tilde{\nabla}\cdot \delta\tilde{W}+\tilde{\sigma}\cdot\delta\tilde{S}\nonumber
\end{equation}
\begin{equation}
=(\delta \rho_e\tilde{E}_0+\delta\tilde{J}\times
\tilde{B}_0)+(\rho_{e0}\delta\tilde{E}+\tilde{J}_0\times
\delta\tilde{B})\ .\label{s1t}
\end{equation}

\section{The static equatorial disk}

We will now restrict our analysis to the investigation of a static
equatorial distribution of matter of thickness $h\lsim r$.
By `static' we mean that the disk fluid is
initially at rest with respect to ZAMOs, i.e. that $v_{0}^i=0$.
Our disk configuration only vaguely mimics astrophysical accretion disks.
We acknowledge that neglecting the Keplerian disk rotation is an important simplification
that we apply only to make some progress with the complex general relativistic
formalism. However, our results may be physically relevant in
Magnetically Arrested Disks in which rotation plays
a secondary role (see discussion section).

The problem we have in
mind is a distribution of matter consisting of two regions inside
and outside some radius $r_0$. We will thus only consider
perturbations in the immediate neighborhood of $r_0$ of the form
\begin{equation}\label{per1}
\left.
\begin{array}{ll}
\delta \rho(t,r,\theta,\phi)=\delta \rho(r)\\
\delta \rho_{e}(t,r,\theta,\phi)=\delta \rho_{e}(r)\\
\delta \upsilon^i(t,r,\theta,\phi)=\delta \upsilon^i(r)\\
\delta B^i(t,r,\theta,\phi)=\delta B^i(r)\\
\delta E^i(t,r,\theta,\phi)=\delta E^i(r)\\
\delta J^i(t,r,\theta,\phi)=\delta J^i(r)
\end{array}
\right\}\cdot e^{nt+i m\phi}
\end{equation}
in the equatorial plane $\theta=\pi/2$, where $m$ takes integer
values $1,2,3$\ldots For simplicity, we will henceforth
ignore the index `0' from the zeroth order terms. In this case,
\begin{eqnarray}\label{l1}
&&v^2=0,~~\Gamma^2=1,~~\delta\Gamma^2=0\nonumber\\
&&\delta S^i=(\rho+p)\delta v^i,~~\delta W^{ij}=\gamma^{ij}\delta p,~~\delta a^\mu=0\nonumber\\
&&\delta\varepsilon=\delta \rho,~~\sigma^{23}=0,~~E^i=0,~~i=r, \theta,\phi\ .
\end{eqnarray}
We will also assume dipolar symmetry in the magnetic field,
namely $B^r(\pi-\theta)=-B^r(\theta)$, $B^\theta(\pi-\theta)=B^\theta(\theta)$, and
$B^\phi(\pi-\theta)=-B^\phi(\theta)$. On the equatorial plane in particular,
\begin{equation}\label{p1}
B^i=(0,B^{\theta},0)\ ,\ a^{\theta}=0\ ,\ \Sigma=r^2\ ,
\end{equation}
and $B_{r,\theta}\sim h_1 B_{\theta}/r$, $B_{\phi,\theta}\sim h_2
B_{\theta}/r$, with $h_1, h_2\approx$~const. In what follows, we will set for
simplicity $h_1=h_2=0$. The system of
eqs.~(\ref{fullMHD1}) \& (\ref{k12}) now becomes
\begin{eqnarray}\label{p2c}
&&\rho_e=0\nonumber\\
&&B^{\theta}=B^{\theta}(r)\nonumber\\
&&J^r=0,~~J^{\theta}=0\nonumber\\
&&J^{\phi}=\frac{\alpha}{4\pi r^2}[B_{\theta,r}+a_r B_{\theta}]\nonumber\\
&&(\rho+p)a^r+(\frac{\Delta}{r^2})
p_{,r}=-\frac{\alpha}{r^2}J_{\phi}B_{\theta}\ .
\end{eqnarray}
The last equation in eqs.~(\ref{p2c}) may be written as
\begin{equation}\label{p2cx}
(p+\frac{B^2}{8\pi})_{,r}=-a_r(\rho+p+\frac{B^2}{4\pi})
+\frac{B^2}{4\pi r}
\end{equation}
To make further progress, we will assume one more simplification, namely
\begin{equation}
\tilde{\nabla}\cdot\delta \tilde{\upsilon}=0\ .\label{anzatz}
\end{equation}

Even under our present assumptions, the general system of first
order equations is rather complicated. We thus decided to move our
detailed calculations to the Appendix B.
Eqs.~(\ref{s1t}) then becomes eq.~(\ref{f1}) which, with the aid of
eqs.~(\ref{y4}) and (\ref{z4}), yields:
\begin{eqnarray}\label{d1}
&&[(\frac{A}{r^4})(r^2\delta\upsilon^r)_{,r}(\rho+p+\frac{B^2}{4\pi})]_{,r}\nonumber\\
&&-a_r(\frac{A}{r^4})(r^2\delta\upsilon^r)_{,r}(\rho+p+\frac{B^2}{4\pi})\nonumber\\
&&=(\frac{m^2}{\Delta})(\rho+p+\frac{B^2}{4\pi})(r^2\delta\upsilon^r)\nonumber\\
&&-\frac{m^2\alpha^2}{n^2+ m^2\omega^2}(\frac{r^2a^r}{A})\{[(1-c_s^4)\rho_{,r}\nonumber\\
&&-(\frac{1}{4\pi})(1+3c_s^2)(\frac{r^2}{2})[(B^{\theta})^2]_{,r}\nonumber\\
&&-(\frac{3}{4\pi})(1+c_s^2)(a_r+\frac{2}{r})B^2]\}(r^2\delta\upsilon^r)\ .
\end{eqnarray}
Eq.~(\ref{d1}) is the general relativistic form of the `force
balance' equation, and is valid inside, outside, and across the
interface $r=r_0$ of two fluids in equilibrium on the equatorial
plane.

In order to make further progress, we will assume that our
physical quantities $\rho$, $p$, and $B$ are constant inside
and outside $r_0$ and $\delta \rho =\delta p
=\delta B =0$ (at least near $r_0$), but may change
discontinuously across $r_0$. $\delta\upsilon^r$ and
$(p+\frac{B^2}{8\pi})$ are continuous\footnote{The continuity of
$\delta\upsilon^r $ is obvious. The continuity of
$(p+\frac{B^2}{8\pi})$ derives from the $r$-derivative terms in
eq.~(\ref{p2cx}).} across the interface between the two fluids,
but $\rho$, $B^2$ and $(\delta\upsilon^r)_{,r}$ in general are
not. For any physical quantity $f$ discontinuous across $r_0$ we
define the jump
\begin{equation}\label{c11y}
{\cal D} \{f\}\equiv f_{(2)}-f_{(1)}\ .
\end{equation}
where
\begin{equation}
f_{(1)}\equiv f(r_0-\epsilon)\ ,\ \mbox{and}\ f_{(2)}\equiv
f(r_0+\epsilon)\ .
\end{equation}
In that notation, eq.~(\ref{d1}) yields
\begin{eqnarray}\label{d1b}
n^2 & = & -m^2 \omega^2\nonumber\\
& + & m^2\left(\frac{r^6 a^r}{A^2}\right)[(1-c_s^4) {\cal
D}\{\rho\}-(1+3c_s^2){\cal D}\{\frac{B^2}{8\pi}\}]\nonumber\\
&& /[{\cal
D}\{(\rho+p+\frac{B^2}{4\pi})(\delta\upsilon^r)_{,r}\}/\delta\upsilon^r]
\label{GeneralD}
\end{eqnarray}

This is the most important equation in our analysis. It is the one
that yields the general criterion for the onset of the magnetic
Rayleight-Taylor instability. The reader can see this directly by
considering the simple un-magnetized Newtonian limit with
$\Delta=r^2$, $A=r^4$, $\omega=a=0$, $\alpha=1$, and $p\ll \rho $.
In that limit, eq.~(\ref{d1b}) yields
\begin{eqnarray}
&&n^2=m^2\frac{a^r {\cal D}\{\rho\}}{{\cal D}\{\rho (\delta
v^r)_{,r}\}/\delta v^r}\ . \label{RTNeutonianx}
\end{eqnarray}
As we will see below, the above denominator
is positive, and therefore, eq.~(\ref{RTNeutonianx}) simply tells
us that the Rayleigh-Taylor instability sets in (i.e. $n^2>0$)
when ${\cal D}\{\rho\}$ has the opposite sign of that of
gravitational acceleration $g^r\equiv -a^r$. The reader can easily
convince him/herself that this indeed corresponds to a `heavy'
fluid above a `light' one (like water over oil). This is
reassuring enough for us to proceed with our investigation. Notice
that $c_s$ is discontinuous across $r_0$, and therefore, the terms
involving $c_s$ in eq.~(\ref{GeneralD}) simply imply average
values across the discontinuity (i.e. $c_s^2\equiv
((c_s)_{(1)}^2+(c_s)_{(1)}^2)/2$ and $c_s^4\equiv
((c_s)_{(1)}^4+(c_s)_{(1)}^4)/2$).

The last missing piece is the calculation of the discontinuity of
$(\delta\upsilon^r)_{,r}$ across $r_0$. This may be obtained by
solving eq.~(\ref{d1}) inside and outside $r_0$ where $\rho$ and
$B^\theta$ are taken to be constant. Eq.~(\ref{d1}) may be
rewritten as
\begin{equation}\label{f19}
(\delta\upsilon^r)_{,rr}+P(r)(\delta\upsilon^r)_{,r}
+Q(r)\delta\upsilon^r=0
\end{equation}
where
\begin{eqnarray}\label{f20}
P(r)&\equiv&\frac{2}{r}+\frac{2}{A}(r^3-a^2 r)-\frac{M}{r(r-2M)}\nonumber\\
Q(r)&=&-\frac{2}{r^2}-\frac{2M}{r^2(r-2M)}+\frac{4(r^3-a^2 M)}{r
A} \nonumber\\
&&-\frac{m^2
r^4}{A\Delta}\nonumber\\
& & -(\frac{\lambda_2}{n^2+ m^2\omega^2})[\frac{M^2 r^2 \Delta
}{A^2(r-2M)^2}]
\end{eqnarray}
where
\begin{equation}\label{f20x}
\lambda_2\equiv m^2
\frac{\frac{3B^2}{4\pi}(1+c_s^2)}{\rho+p+\frac{B^2}{4\pi}}
\end{equation}
In eq.~(\ref{f19}) setting
\begin{eqnarray}\label{f21}
\delta\upsilon^r (r)&=&z(r)\exp{[-\frac{1}{2}\int^r P(t) dt]}\nonumber\\
&&=z(r)\sqrt{\frac{\alpha}{A}}
\end{eqnarray}
where
\begin{equation}\label{pr1}
I=\frac{1}{2}\int^r P(t)
dt=\frac{1}{4}\ln{r}-\frac{1}{4}\ln{\Delta}+\frac{3}{4}\ln{\frac{A}{r}}
\end{equation}
we obtain a simpler form of eq.~(\ref{f19}), namely
\begin{equation}\label{f22}
\frac{d^2 z(r)}{dr^2}+q(r) z(r)=0 \label{harmonic}
\end{equation}
with
\begin{eqnarray}\label{f22b}
&&q(r)\equiv Q(r)-\frac{1}{2}\frac{d P(r)}{dr}-\frac{1}{4}P(r)^2\nonumber\\
&&=-\frac{M(8r-11M)}{4r^2(r-2M)^2}+\frac{Mr^2}{A(r-2M)}-\frac{2}{r^2}-\frac{m^2 r^4}{A\Delta}\nonumber\\
&&+\frac{r^2(3r^4-A)}{A^2}-\frac{\lambda_2 M^2 r^2 \Delta}{(n^2+m^2\omega^2)A^2(r-2M)^2}\nonumber\\
&&+\frac{a^2}{A^2}(2r^4-2a^2M r-3 a^2 M^2)\nonumber\\
&&-\frac{a^2 M}{rA(r-2M)}(2r-3M)\ .\nonumber\\
\end{eqnarray}
Eq.~(\ref{harmonic}) is reminiscent of the equation of a harmonic
oscillator. Obviously, we do not plan to solve the general form of
this equation, since after all we are interested only in what
happens around our reference radius $r_0$. We will thus consider
next particular limiting cases.


\section{The Schwarzschild Case}

\subsection{Un-magnetized}

We first consider the un-magnetized non-rotating case with
$a=\omega=B^2=0$. In this case $A=r^4$, $\Delta =r^2-2 M r$, and
eq.~(\ref{d1}) simplifies considerably while eq.~(\ref{d1b}) becomes

\begin{equation}
n^2=\frac{m^2 a^r (1-c_s^4){\cal D}\{\rho\}}{r^2{\cal
D}\{(\rho+p)(\delta\upsilon^r)_{,r}\}/\delta\upsilon^r}\label{c2}
\end{equation}

Taking into account the above considerations, eq.~(\ref{f22})
admits the general solution
\[
z(r)=c_2 r^{(1/4)}\sqrt{(r-2M)}P_{\xi-1/2}^{3/2}(\sqrt{\frac{r}{2M}})
\]
\begin{equation}\label{c7}
+c_1 r^{(1/4)}\sqrt{(r-2M)}Q_{\xi-1/2}^{3/2}(\sqrt{\frac{r}{2M}})
\end{equation}
where $c_1 ,c_2$ are arbitrary constants,
$\xi\equiv\sqrt{1+4 m^2}$, and
$P_{\xi-1/2}^{3/2}(\sqrt{\frac{r}{2M}})$ and
$Q_{\xi-1/2}^{3/2}(\sqrt{\frac{r}{2M}})$ are the Legendre
associate functions of first and second order respectively. In the
limit $M\rightarrow 0$ these functions behave as
\begin{eqnarray}\label{c8}
&&P_{\nu}^{\mu}(z)\sim \frac{\Gamma(\nu+1/2)}{\sqrt{\pi}\Gamma(\nu+\mu+1)} (2z)^{\nu},\nonumber\\
&&Q_{\nu}^{\mu}(z)\sim\frac{\sqrt{\pi}}{2^{\nu+1}\Gamma(\nu+3/2)
z^{\nu+1}}
\end{eqnarray}
(Oliver~1974), where $z=\sqrt{\frac{r}{2M}}$,
$\nu=\xi-\frac{1}{2}$ and $\mu=\frac{3}{2}$. Furthermore, we
define
\begin{eqnarray}\label{y1}
&&\delta\upsilon_{(1)}^r(r)=c_1 F(r)P_{\xi-1/2}^{3/2}(\sqrt{\frac{r}{2M}})e^{nt+i m\phi}\nonumber\\
&&\sim c_1 A_1 (1-\frac{2M}{r})^{3/4} r^{\xi/2-3/2}e^{nt+ i m\phi},~~\mbox{~for~}~~r<r_0\nonumber\\
&&\delta\upsilon_{(2)}^r(r)=c_2 F(r)Q_{\xi-1/2}^{3/2}(\sqrt{\frac{r}{2M}})e^{nt+i m\phi}\nonumber\\
&&\sim c_2 A_2 (1-\frac{2M}{r})^{3/4} r^{-\xi/2-3/2}e^{nt+ i m\phi},~~\mbox{~for~}~~r>r_0\nonumber\\
\end{eqnarray}
where $F(r)\equiv[\frac{2^{1/4}(r-2M)^{3/4}}{r^2}]$ and
\begin{eqnarray}\label{y3}
&&A_1\equiv\frac{2^{\xi/2-1/4}}{\sqrt{\pi} M^{(\xi/2-1/4)} \xi(1+\xi)}=
\mbox{const.}\nonumber\\
&&A_2\equiv\frac{\sqrt{\pi}
M^{(\xi/2+1/4)}}{2^{\xi/2+1/4}\xi\Gamma(\xi)} =\mbox{const.}
\end{eqnarray}

Putting everything back into eq.~(\ref{c2}), after long
calculations, we obtain the simple result
\begin{equation}\label{t5x}
n^2=\frac{m^2 (\frac{a^r}{r}) (1-\frac{2M}{r}) (1-c_s^4){\cal D}
\{\rho\}}{\xi(\rho+p)(1-\frac{2M}{r})-\frac{1}{2}(1+\frac{M}{r}){\cal
D}\{\rho\}}\ .
\end{equation}
Notice that all quantities that appear in the above equation imply
their averages across the interface $r=r_0$ (e.g. $\rho\equiv
(\rho_{(1)}+\rho_{(2)})/2$, etc). For $M=0$, eq.~(\ref{t5x})
reduces to eq.~(51) in Chap.~X of Chandrasekhar~(1961).

\subsection{Magnetized}
The `force-balance' eq.~(\ref{d1}) becomes very complicated in the
general magnetized case. In what follows, we will consider the
general form of the stability criterion (eq.~\ref{d1b}), but will
at the same time adopt the expressions for $\delta v^r$ across the
interface that we obtained in the unmagnetized case. Under this
approximation eq.~(\ref{d1b}) yields
\[
n^2 = m^2 (\frac{a^r}{r}) (1-\frac{2M}{r}) \{ (1-c_s^4){\cal D}
\{\rho\}-(1+3c_s^2){\cal D}\{\frac{B^2}{8\pi}\}]\}\]
\begin{equation}
/ [\xi(\rho+p+\frac{B^2}{4\pi})(1-\frac{2M}{r})
-\frac{1}{2}(1+\frac{M}{r}){\cal D}\{\rho+p+\frac{B^2}{4\pi}\}]
\label{magn1}\end{equation}
The denominator of the above
equation is always positive, thus the sign of $n^2$ is dictated by
the sign of the numerator. Thus, in the limit $c_s\rightarrow 0$,
the criterion for instability in the magnetized Schwarzschild case
becomes
\begin{equation}\label{magn2}
{\cal D}\{\rho\}-{\cal D}\{\frac{B^2}{8\pi}\}>0\ ,
\end{equation}
which is the same as eq.~(234) in Chap.~X of Chandrasekhar~(1961)
obtained in the Newtonian limit.


\section{The Kerr Case}
In Sec.4, we have examined the Rayleigh-Taylor instability in the
presence of a dynamically significant magnetic field in a
Schwarzschild space-time using the approximation that the
solutions for $\delta v^r$ inside and outside the interface $r=r_0$
are those obtained in the un-magnetized Schwarzschild case. We
will apply a similar approximation in the Kerr case. The
`force-balance' equation (eq.~\ref{z5}) now becomes complex and
results in two independent equations on the interface
(eqs.~\ref{f1} and \ref{f2}). In what follows, we will consider only
the first equation, $\Lambda_1=(N_1)_{,r}$, since the second
equation (\ref{second}) yields a similar stability criterion. The expressions for
$\Lambda_1$ and $N_1$ are given in eqs.~(\ref{y4}) and (\ref{y5})
in the Appendix B.

As we pointed out above, we will proceed using the solutions of
eq.~(\ref{f22}) with $\lambda_2=0$ as in the un-magnetized
Schwarzschild case, only now $a\ne 0$. Because of its complexity, eq.~(\ref{f22b}) is
still rather difficult to be solved analytically. However, if we
only consider slowly rotating Kerr black holes with relatively
small $a$, we can expand (\ref{f22b}) in powers of $a$ and keep
terms up to $a^2$. Next, we expand the coefficient of $a^2$ in powers
of $1/r$, and keep only terms up to $1/r$ and $1/(r-2M)$.
In this case, eq.~(\ref{f22b}) becomes
\begin{equation}\label{new1}
\frac{d^2 z(r)}{dr}=-[q_{S}(r)+a^2 q_{\rm K}(r)]z(r)
\end{equation}
where $q_S(r)$, $q_{\rm K}(r)$ correspond to the Schwarzschild and Kerr space-times,
respectively and their explicit forms are
\begin{eqnarray}\label{new2}
&&q_S(r)\equiv\frac{3M^2-4Mr-4 m^2(r^2-2Mr)}{4r^2(r-2M)^2}\nonumber\\
&&q_{\rm K}(r)\equiv\frac{8m^4-18m^2+9}{32 m^2 M^2(r-2M)}-\frac{8m^4-18m^2+9}{32 m^2 M^3 r}
\end{eqnarray}
Eq.~(\ref{new1}) admits two general solutions
\begin{eqnarray}\label{Kerr2}
z_1(r)&=&c_1 r^{1/4}(1-\frac{2M}{r})^{1/2} P_{\xi_{\rm K}-1/2}^{3/2}(\sqrt{\frac{r}{2M}})\nonumber\\
z_2(r)&=&c_2 r^{1/4}(1-\frac{2M}{r})^{1/2} Q_{\xi_{\rm K}-1/2}^{3/2}(\sqrt{\frac{r}{2M}})\nonumber\\
\end{eqnarray}
where $\xi_{\rm K}=\sqrt{(1+4m^2)-\frac{a^2}{4M^2 m^2}(m^2-\frac{3}{4})(m^2-\frac{3}{2})}$.

Observe, that the solutions (\ref{Kerr2}) differ from those of eqs.~(\ref{c7}) only in
the indices. Namely, in the Schwarzschild case, index $\xi_{\rm K}$ becomes equal to
$\xi$. All the other factors in eqs.~(\ref{Kerr2}) are the same as in eq.~(\ref{c7}).
Thus, following the computations of subsection (4.1) and keeping terms only up to
second order in $a$ we end up with the criterion
\begin{eqnarray}\label{Kerr9}
n^2 & = & -m^2\omega^2\nonumber\\
& + & m^2(\frac{M}{r^3})\frac{r^6 \Delta}{A^2}
[(1-c_s^4){\cal
D}\{\rho\}
-(1+3c_s^2){\cal D}\{\frac{B^2}{8\pi}\}]\nonumber\\
& / & \{[\frac{r-2M}{2\Delta A}[(6r^3-4Mr^2)a^2
+(4r-9M)r^4]\nonumber\\
&& +\frac{4M}{r}-\frac{5}{2}]{\cal D}\{\rho+p+\frac{B^2}{4\pi}\}\nonumber\\
&&+\xi_{\rm K}(1-\frac{2M}{r})(\rho+p+\frac{B^2}{4\pi})\}
\end{eqnarray}
One can easily verify that when $a^2=0$, eq.~(\ref{Kerr9}) reduces to eq.~(\ref{magn1}).

As before, it is easy to show that the denominator in the r.h.s. of eq.~(\ref{Kerr9}) is
always positive, and thus the stability criterion depends on the sign of the
numerator. However, the new element here is that the Kerr geometry introduces a new term
in the r.h.s., namely $-m^2\omega^2$ which softens the instability criterion.
Thus, a configuration which would have been unstable in a non-rotating space-time, may
now become stable. In other words, {\em the rotation of the space-time works in
a direction that reduces the Rayleigh-Taylor instability}. As we will see next, this unexpected result has very important astrophysical applications.


\section{Summary and Discusssion}

Our goal has been to obtain a criterion for the onset of the magnetic Rayleight-Taylor instability in curved space time. In order to achieve this goal, we made the following simplifying idealized assumptions:
\begin{enumerate}
\item We considered a disk configuration stationary with respect to ZAMOs (i.e. with velocity given by eq.~\ref{k3x}).
\item We investigated only what happens on the equatorial plane $\theta=\pi/2$, and in particular in the vicinity of some interface at radius $r=r_0$.
\item We assumed dipolar symmetry in the magnetic field. The latter resulted in $B^i=(0,B^{\theta},0)$, and $J^{\mu}=(0,0,0,J^{\phi})$.
\item We assumed ideal MHD conditions in the form of eq.~(\ref{l1x}).
\item In order to make further progress, we assumed that $\tilde{\nabla}\cdot \tilde{\delta \upsilon}=0$, and that $\rho$, $p$, and $B$ are uniform throught the disk, with the exception of a jump in their values at some interface $r_0$.
\end{enumerate}

Under the above conditions, we perturbed all physical quantities appearing in eqs.~(3)-(9) to first order, we obtained the zero and first order equations (eqs.~73 and 74-85 respectively) under the Cowling approximation $\delta g_{\mu\nu}=0$, and ended up with eq.~(86) in the complex plane. The real part of that equation, eq.~(35), is used to obtain both the dependence of the unknown function $\delta\upsilon^r$ on the radial coordinate $r$ away from the interface $r=r_0$, and the stability criterion at the interface itself. Notice that $\delta\upsilon^r$ and $(p+\frac{B^2}{8\pi})$ are continuous across the interface, but $\rho$, $B^2$ and $(\delta\upsilon^r)_{,r}$ in general are not.

Eq.~(52) expresses the stability criterion in the un-magnetized Schwarzschild space-time. To obtain the criterion in the magnetized Schwarzschild case, eq.~(53), we used the solution for $\delta\upsilon^r$  obtained in the unmagnetized case, eq.~(45). Similarly, to obtain the criterion in the magnetized Kerr case, eq.~(58), we used the solution for $\delta\upsilon^r$ obtained in the unmagnetized case, eq.~(55).

\subsection{Astrophysical Implications}

Let us here obtain a crude estimate of the maximum value of the magnetic field for which the disk-field configuration is stable. This is roughly also the maximum value of the magnetic field that can be held inside
the inner edge of the disk at some radius $r_0$ around the innermost stable orbit (ISCO) of a spinning black hole. In the limit of small $a^2$, negligible magnetic and gas pressure compared to the rest mass energy density $\rho$,\footnote{A crude estimate of the rest mass energy density at the Eddington accretion rate is $\rho\sim GM m_p/r_0^2 \sigma_T\sim 4\times 10^{14}M_1^{-1}$~erg/cm$^3$, where $M_1$ is the mass of the black hole in solar mass units, and $\sigma_T$ is the electron Thomson cross section. The magnetic field $B$ must be well below its equipartition value of $B_{\rm eq}\sim 10^8 M_1^{-1/2}$~G (eq.~2) for our assumption of neglible magnetic pressure to apply.} and assuming a continuous matter distribution ${\cal D}\{\rho\}=0$ through the interface, eq.~(\ref{Kerr9}) yields the stability criterion (in real units)
\begin{equation}
 -{\cal D}\{\frac{B^2}{8\pi}\}\lsim
\frac{\omega^2}{\Omega_{\rm K}^2}\rho \approx
\left(\frac{r_0}{r_{\rm S}}\right)^{-3}\left(\frac{a}{M}\right)^2\rho
\label{simple1}
\end{equation}
(factors of order unity have been dropped from this calculation). $r_{\rm S}=2GM/c^2$ is the Schwarzschild radius, and $\Omega_{\rm K}^2\equiv GM/r_0^3$.
We have considered here only the most unstable mode with $m=1$ with $\xi_{\rm K}\approx 2$, and assumed that $r_0\sim 6GM/c^2$. If we further assume for simplicity that ${\cal D}\{B^2\}\approx -B^2$, i.e. if we assume that most of the field is brought inside $r_0$, eq.~(\ref{simple1}) yields
\begin{equation}
\frac{B_{\rm max}^2}{8\pi} \sim \frac{GM M_{\rm disk}}{4\pi r_0^4}\left(\frac{a}{M}\right)^2\ ,
\label{criterion}
\end{equation}
which differs from the result of eq.~(\ref{2}) by a factor of order $(a/M)^2$! We have
assumed here a thick disk with mass $M_{\rm disk}\sim 4\pi r_0^3 \rho/c^2$, and, as before, factors of order unity have been dropped from this order of magnitude estimate. The calculation may be crude, but leads to an important result, namely that even a small amount of poloidal magnetic field held inside the inner edge of an astrophysical accretion disk is unstable to the magnetic Rayleigh-Taylor instability, {\em unless} the central black hole is spinning.

One implication of this result is that non-rotating Magnetically Arrested Disks cannot exist around non-rotating black holes. MADs were first obtained in 2D general relativistic simulations where the Rayleigh-Taylor instability is obviously absent (e.g. Tchekhovskoy et al.~2010). MADs have also been obtained in 3D non-relativistic MHD simulations (e.g. Igumenshchev et al.~2003; Narayan et al.~2003) where rotation may play an important role in stabilizing the innermost disk against the Rayleigh-Taylor instability. Notice that Tchekhovskoy et al.~(2012) sampled the full range of $a/M$ and didn't observe the decrease in the average flux accuulated through the black hole horizon for low black hole spins implied by our present results\footnote{In fact, they observed a slight decrease at high black hole spins which we believe may be associated to the shrinking of the black hole horizon with spin.}. We can only speculate that this is due to accretion: magnetic flux is advected inwards and at the same time escapes due to Rayleigh-Taylor instability, thus, on average, the amount of accumulated magnetic flux is non-zero. We may be able to account for the effect of accretion in a future publication.


We conclude by emphasizing that the magnetic Rayleigh-Taylor instability has serious implications for the origin of astrophysical jets and their associated radio emission. It is generally considered that some amount of the magnetic flux that is held by the accretion disk threads the horizon of the central black hole. As a result, relativistic jet outflows are expected both from the vicinity of the black hole and the inner accretion disk, therefore, it is hard to separate their respective contributions to the total jet power (Christodoulou et al,~2016, submitted). Observations tend to support such a combined structure with the corresponding models referred to as ``spine - sheath" (Ghisellini et al~2005), with both components contributing to the jet power. According to Blandford \& Znajek~(1977), if the central black hole is spinning, a highly relativistic black hole jet is generated which extracts power
\begin{equation}
P_{\rm BZ} \sim \frac{1}{c}B_{\rm BH}^2r_{\rm BH}^4 \omega_{\rm BH}^2
\end{equation}
(Blandford \& Znajek~1977; Tchekhovskoy et al.~2010; Nathanail \& Contopoulos~2014). Here, $B_{\rm BH}$ is the value of the magnetic field that threads the black hole horizon (this is roughly the same as the value of the magnetic field that is held inside the inner edge of the disk), $r_{\rm BH}$ is the radius of the horizon, and $\omega_{\rm BH}$ is the black hole angular frequency. It is, therefore, imperative to understand  how the magnetic Rayleigh-Taylor instability limits the maximum possible accumulated magnetic flux. We thus plan to continue our investigation in the presence of accretion and rotation.

\section*{Acknowledgements}
This work was supported by the General Secretariat for Research
and Technology of Greece and the European Social Fund in the
framework of Action `Excellence'.

\section*{Appendix A: Useful expressions}
Below we have collected some useful expressions
\begin{eqnarray}\label{m1}
\frac{1}{\alpha}[\tilde{\nabla}\cdot (\alpha \tilde{J}]&=&\tilde{\nabla}\cdot \tilde{J}+\tilde{J}\cdot \frac{\tilde{\nabla}\alpha}{\alpha}=\tilde{\nabla}\cdot \tilde{J}+\tilde{J}\cdot \tilde{a}\nonumber\\
&&=[J_{,i}^i+\Gamma_{il}^i J^l]+\gamma_{ij} a^i J^j\ ,
\end{eqnarray}
where $\tilde{a}=\tilde{\nabla}\alpha/\alpha$ in the acceleration,
\begin{eqnarray}
\frac{1}{\alpha^2}[\tilde{\nabla}\cdot (\alpha^2 \tilde{S}]&=&\tilde{\nabla}\cdot \tilde{S}+\tilde{S}\cdot \frac{\tilde{\nabla}\alpha^2}{\alpha^2}=\tilde{\nabla}\cdot \tilde{S}+2\tilde{J}\cdot \tilde{a}\nonumber\\
&&=[S_{,i}^i+\Gamma_{il}^i S^l]+\gamma_{ij} a^i S^j\nonumber\\
\frac{1}{\alpha}[\tilde{\nabla}\cdot (\alpha \tilde{W}]&=&\tilde{\nabla}\cdot \tilde{W}+\tilde{W}\cdot \frac{\tilde{\nabla}\alpha}{\alpha}=\tilde{\nabla}\cdot \tilde{W}+\tilde{W}\cdot \tilde{a}\nonumber\\
&&=[W_{,i}^{ki}+\Gamma_{il}^i W^{kl}]+\gamma_{ij} a^i W^{kj}\\
\frac{1}{\alpha}[\tilde{\nabla}\times (\alpha \tilde{B}]&=&\tilde{\nabla}\times \tilde{B}-\tilde{B}\times\frac{\tilde{\nabla}\alpha}{\alpha}=\tilde{\nabla}\times \tilde{B}-\tilde{B}\times \tilde{a}\nonumber\\
\frac{1}{\alpha}[\tilde{\nabla}\times (\alpha
\tilde{E}]&=&\tilde{\nabla}\times
\tilde{E}-\tilde{E}\times\frac{\tilde{\nabla}\alpha}{\alpha}=\tilde{\nabla}\times
\tilde{E}-\tilde{E}\times \tilde{a}
\label{m2}\end{eqnarray} In
Kerr space-time
\begin{equation}\label{m3}
\frac{1}{\alpha}[\rho_{,\mu}(\alpha
U^{\mu}+\beta^{\mu})-\gamma^{ij}\beta_i
\rho_{,j}]=\frac{1}{\alpha}[\rho_{,0}+\omega\rho_{,\phi}]
\end{equation}

\section*{Appendix B: GRMD equations}

Below we summarize the general relativistic MHD equations.

\subsection{Zeroth Order Equations}

\begin{eqnarray}\label{p2}
&&\rho_e=0\nonumber\\
&&B_{,\theta}^{\theta}+\frac{\cot{\theta}}{\Sigma}[r^2+a^2M^2(1-3\sin^2{\theta})]B^{\theta}=0\,~~B_{,\phi}^{\theta}=0\nonumber\\
&&J^r=-\frac{\alpha}{4\pi\Sigma\sin{\theta}}B_{\theta,\phi},~~J^{\theta}=0\nonumber\\
&&J^{\phi}=\frac{\alpha}{4\pi\Sigma\sin{\theta}}[B_{\theta,r}-B_{r,\theta}+a_r B_{\theta}]\nonumber\\
&&J_{,r}^r+J_{,\phi}^{\phi}+(\frac{\Sigma}{\Delta}a^r+\frac{2r}{\Sigma})J^r=0\nonumber\\
&&\rho_{,\phi}=0\Rightarrow \rho=\rho(r,\theta)\nonumber\\
&&(\rho+p)a^r+(\frac{\Delta}{\Sigma})p_{,r}=-\frac{\alpha}{\Sigma\sin{\theta}}J_{\phi}B_{\theta}\nonumber\\
&&(\rho+p)a^{\theta}+(\frac{1}{\Sigma})p_{,\theta}=0\nonumber\\
&&\frac{A}{\Sigma\sin^2{\theta}}p_{,\phi}
=\frac{\alpha}{\Sigma\sin{\theta}}J_r B_{\theta}\ .
\end{eqnarray}

\subsection{First Order Equations}

\begin{eqnarray}\label{p3}
&&\delta E^r=\frac{\alpha}{\Sigma\sin{\theta}}B_{\theta}\delta \upsilon_{\phi},~~\delta E^{\theta}=0,~~\delta E^{\phi}=-\frac{\alpha}{\Sigma\sin{\theta}}B_{\theta}\delta \upsilon_r\nonumber\\
&&\delta E_{,r}^r+\delta E_{,\phi}^{\phi}+\frac{1}{\Sigma}[2r\delta E^r+(\cot{\theta}\Sigma -2a^2M^2\sin{\theta}\cos{\theta})\delta E^{\theta}]\nonumber\\
&&=4\pi\delta \rho_e\nonumber\\
&&\delta B_{,r}^r+\delta B_{,\theta}^{\theta}+\delta B_{,\phi}^{\phi}\nonumber\\
&&+\frac{1}{\Sigma}[2r\delta B^r+(\cot{\theta}\Sigma -2a^2M^2\sin{\theta}\cos{\theta})\delta B^{\theta}]=0\nonumber\\
\end{eqnarray}
\begin{eqnarray}\label{p4}
&&4\pi\delta J^r=\frac{\alpha}{\Sigma\sin{\theta}}[\delta B_{\phi,\theta}-\delta B_{\theta,\phi}+a_{\theta}\delta B_{\phi}]\nonumber\\
&&-\frac{1}{\alpha}[\delta E_{,t}^r+\omega\delta E_{,\phi}^r]-\frac{\delta E^{\phi}}{\alpha}(\Gamma_{t\phi}^r+\omega\Gamma_{\phi\phi}^r)+\gamma_{\phi\phi}\sigma^{r\phi}\delta E^{\phi}\nonumber\\
&&4\pi\delta J^{\theta}=\frac{\alpha}{\Sigma\sin{\theta}}[\delta B_{r,\phi}\nonumber\\
&&-\delta B_{\phi,r}-a_r\delta B_{\phi}]-\frac{\delta E^{\phi}}{\alpha}(\Gamma_{t\phi}^{\theta}+\omega\Gamma_{\phi\phi}^{\theta})\nonumber\\
&&4\pi\delta J^{\phi}=\frac{\alpha}{\Sigma\sin{\theta}}[\delta B_{\theta,r}-\delta B_{r,\theta}+a_r\delta B_{\theta}-a_{\theta}\delta B_r]\nonumber\\
&&-\frac{1}{\alpha}[\delta E_{,t}^\phi+\omega\delta E_{,\phi}^\phi]-\frac{\delta E^r}{\alpha}(\Gamma_{t r}^\phi+\omega\Gamma_{\phi r}^\phi-\omega a_r)\nonumber\\
&&+\gamma_{rr}\sigma^{\phi r}\delta E^r
\end{eqnarray}
\begin{eqnarray}\label{p5}
&&\frac{1}{\alpha}(\delta B_{,t}^r+\omega\delta B_{,\phi}^r)+(\Gamma_{t\phi}^r+\omega\Gamma_{\phi\phi}^r)\frac{\delta B^{\phi}}{\alpha}-\gamma_{\phi\phi}\sigma^{r\phi}\delta B^{\phi}\nonumber\\
&&=\frac{\alpha}{\Sigma\sin{\theta}}(\delta E_{\theta,\phi}-\delta E_{\phi,\theta}-a_{\theta}\delta E_{\phi})\nonumber\\
&&\frac{1}{\alpha}(\delta B_{,t}^{\theta}+\omega\delta B_{,\phi}^{\theta})+(\Gamma_{t\phi}^{\theta}+\omega\Gamma_{\phi\phi}^{\theta})\frac{\delta B^{\phi}}{\alpha}-\gamma_{\phi\phi}\sigma^{\theta\phi}\delta B^{\phi}\nonumber\\
&&=\frac{\alpha}{\Sigma\sin{\theta}}(\delta E_{\phi,r}-\delta E_{r,\phi}+a_{r}\delta E_{\phi})\nonumber\\
&&\frac{1}{\alpha}(\delta B_{,t}^{\phi}+\omega\delta B_{,\phi}^{\phi})+(\Gamma_{t r}^{\phi}+\omega\Gamma_{\phi r}^{\phi}-\omega a_r)\frac{\delta B^{r}}{\alpha}\nonumber\\
&&+(\Gamma_{t \theta}^{\phi}+\omega\Gamma_{\phi \theta}^{\phi}-\omega a_\theta)\frac{\delta B^{\theta}}{\alpha}-\gamma_{rr}\sigma^{\phi r}\delta B^r-\gamma_{\theta\theta}\sigma^{\phi\theta}\delta\upsilon^{\theta}\nonumber\\
&&=\frac{\alpha}{\Sigma\sin{\theta}}(\delta E_{r,\theta}-\delta E_{\theta,r}-a_{r}\delta E_{\theta}+a_{\theta}\delta E_r)\nonumber\\
\end{eqnarray}
\begin{eqnarray}\label{p6}
&&\frac{1}{\alpha}(\delta \rho_{e,t}+\omega\delta \rho_{e,\phi})=-(\frac{\Sigma}{\Delta})a^r \delta J^r-\Sigma a^{\theta} \delta J^{\theta}\nonumber\\
&&+\delta J_{,r}^r+\delta J_{,\theta}^{\theta}+\delta J_{,\phi}^{\phi}\nonumber\\
&&+\frac{1}{\Sigma}[2r \delta J^r+(\Sigma \cot{\theta}-2a^2
m^2\sin{\theta}\cos{\theta})\delta J^{\theta}]
\end{eqnarray}
\begin{eqnarray}\label{p7}
&&\frac{1}{\alpha}(\delta \rho_{,t}+\omega\delta \rho_{,\phi})+2(\rho+p)[a_r\delta \upsilon^r+ a_{\theta}\delta \upsilon^{\theta}]\nonumber\\
&&+[\delta \upsilon^r (\rho+p)_{,r}+\delta \upsilon^{\theta}(\rho+p)_{,\theta}+\delta \upsilon^{\phi}(\rho+p)_{,\phi}]\nonumber\\
&&=-(\frac{\Sigma}{\Delta}) J^r\delta E^r-(\frac{A\sin^2{\theta}}{\Sigma})J^{\phi}\delta E^{\phi}\nonumber\\
\end{eqnarray}

\begin{eqnarray}\label{p8}
&&\frac{(\rho+p)}{\alpha}(\delta \upsilon_{,t}^r+\omega\delta \upsilon_{,\phi}^r)+\frac{(\rho+p)}{\alpha}(\Gamma_{t\phi}^r+\omega\Gamma_{\phi\phi}^r)\delta\upsilon^{\phi}\nonumber\\
&&+a^r(\delta\rho+\delta p)+(\frac{\Delta}{\Sigma}) \delta p_{,r}+(\rho+p)\gamma_{\phi\phi}\sigma^{r\phi}\delta \upsilon^{\phi}\nonumber\\
&&=-\frac{\alpha}{\Sigma\sin{\theta}}(\delta J_{\phi} B_{\theta}-\delta B_{\phi} J_{\theta}+J_{\phi}\delta B_{\theta})\nonumber\\
&&\frac{(\rho+p)}{\alpha}(\delta \upsilon_{,t}^{\theta}+\omega\delta \upsilon_{,\phi}^{\theta})+\frac{(\rho+p)}{\alpha}(\Gamma_{t\phi}^{\theta}+\omega\Gamma_{\phi\phi}^{\theta})\delta\upsilon^{\phi}\nonumber\\
&&+a^{\theta}r(\delta\rho+\delta p)+(\frac{1}{\Sigma}) \delta p_{,\theta}+(\rho+p)\gamma_{\phi\phi}\sigma^{\theta\phi}\delta\upsilon^{\phi}\nonumber\\
&&=-\frac{\alpha}{\Sigma\sin{\theta}}(J_{r} \delta B_{\phi}-\delta B_{r} J_{\phi})\nonumber\\
&&\frac{(\rho+p)}{\alpha}(\delta \upsilon_{,t}^{\phi}+\omega\delta \upsilon_{,\phi}^{\phi})+\frac{(\rho+p)}{\alpha}[(\Gamma_{t r}^{\phi}+\omega\Gamma_{\phi r}^{\phi}-\omega a_r)\delta\upsilon^{r}\nonumber\\
&&+(\Gamma_{t\theta}^{\phi}+\omega\Gamma_{\phi \theta}^{\phi}-\omega a_\theta)\delta\upsilon^{\theta}]+(\frac{\Sigma}{A\sin^2{\theta}}) \delta p_{,\phi}\nonumber\\
&&+(\rho+p)(\gamma_{rr}\sigma^{\phi r}\delta\upsilon^r+\gamma_{\theta\theta}\sigma^{\phi\theta}\delta\upsilon^{\theta})\nonumber\\
&&=-\frac{\alpha}{\Sigma\sin{\theta}}(\delta B_{r} J_{\theta}-\delta J_{r} B_{\theta}-J_r\delta B_{\theta})\nonumber\\
\end{eqnarray}

\subsection{The equatorial plane}
On the equatorial plane the zero order eqs.~(\ref{p2}), with contravariant and some of them
with covariant indices needed for our work, reads
\begin{eqnarray}\label{x1}
&&\rho_e=0\nonumber\\
&&B_{,\theta}^{\theta}=0,~~B_{,\phi}^{\theta}=0\nonumber\\
&&J^r=0,~~J^{\theta}=0,~~J_{,\phi}^{\phi}=0\nonumber\\
&&J^{\phi}=\frac{\alpha}{4\pi\Sigma}[B_{\theta,r}+a_r B_{\theta}]\nonumber\\
&&J_r=0,~~J_{\theta}=0,\nonumber\\
&&J_{\phi}=(\frac{1}{4\pi})(\frac{\Delta}{\alpha})(B_{,r}^{\theta}+\frac{2}{r}B^{\theta}+a_r B^{\theta}),\nonumber\\
&&\rho_{,\phi}=0\Rightarrow \rho=\rho(r,\theta)\nonumber\\
&&(\rho+p)a^r+(\frac{\Delta}{\Sigma})p_{,r}=-\frac{\alpha}{\Sigma} J_{\phi}B_{\theta}\nonumber\\
&&(\rho+p)a^{\theta}+(\frac{1}{\Sigma})p_{,\theta}=0\nonumber\\
&& p_{,\phi} =\frac{\alpha}{A} J_r B_{\theta}\ .
\end{eqnarray}

The first order equations eqs.~(\ref{p3})-(\ref{p8}) simplifies considerably on the equatorial
plane, where we have $a^{\theta}=0$, $\sigma^{\theta\phi}=0$, $\Gamma_{t\phi}^{\theta}=\Gamma_{\phi\phi}^{\theta}=0$, and $\Gamma_{t\theta}^{\phi}=\Gamma_{\theta\phi}^{\phi}=0$. Our anzatz (eq.~\ref{anzatz}), because of eqs.~(\ref{per1})
becomes
\begin{eqnarray}\label{inc2}
&&\delta \upsilon_{,r}^r+\delta \upsilon_{,\theta}^{\theta}+\delta
\upsilon_{,\phi}^{\phi}+\frac{2}{r}\delta \upsilon^r=0
\end{eqnarray}
which in turn, gives
\begin{equation}\label{ff1}
- i m\delta \upsilon^{\phi}=\frac{1}{r^2}(r^2 \delta
\upsilon^r)_{,r}\equiv\chi\ .
\end{equation}
Furthermore, with the aid of eqs.~(\ref{p2c}), the system of first
order equations, eqs.~(\ref{p3})-(\ref{p8}) and (\ref{ff1}), yield
\begin{eqnarray}\label{p3b}
&&\delta B^{r}=\delta B^{\phi}=0\nonumber\\
&&\delta B^{\theta}=-(n-i m \omega)\alpha\Xi_1\delta\upsilon^r
\end{eqnarray}
where $\Xi_1=\frac{B^{\theta}_{,r}}{n^2+ m^2\omega^2}$.\newline\

Eqs.~(\ref{p3}) yield
\begin{eqnarray}\label{p3a}
&&\delta E^r=\frac{\alpha}{r^2}B_{\theta}\delta \upsilon_{\phi},~~\delta E^{\theta}=0,~~\delta E^{\phi}=-\frac{\alpha}{r^2}B_{\theta}\delta \upsilon_r\nonumber\\
&&\delta E_r=\frac{r^2}{\alpha}B^{\theta}\delta \upsilon^{\phi},~~\delta E_{\theta}=0,~~\delta E_{\phi}=-\frac{r^2}{\alpha}B^{\theta}\delta \upsilon^r\nonumber\\
&&\delta E_{,r}^r+\delta E_{,\phi}^{\phi}+\frac{2}{r}\delta E^r=4\pi\delta \rho_e\ .
\end{eqnarray}
Eq.~(\ref{p7}) gives
\begin{eqnarray}\label{r2}
&&\delta\rho=-\frac{\alpha(n-i m\omega)}{n^2+
m^2\omega^2}[(1-c_s^2)\rho_{,r}\nonumber\\
&&-\frac{3}{4\pi}(B_{\theta}B_{,r}^{\theta}+\frac{2r}{\Sigma} B^2 +a_r B^2)]\delta \upsilon^r
\end{eqnarray}
Eqs.~(\ref{p4}) become
\begin{eqnarray}\label{p5b}
&&4\pi\delta J^r=-\frac{\alpha}{\Sigma}\delta B_{\theta,\phi}\nonumber\\
&&-\frac{1}{\alpha}[\delta E_{,t}^r+\omega\delta E_{,\phi}^r]-\frac{\delta E^{\phi}}{\alpha}(\Gamma_{t\phi}^r+\omega\Gamma_{\phi\phi}^r-\alpha\gamma_{\phi\phi}\sigma^{r\phi}),\nonumber\\
&&4\pi\delta J^{\theta}=0,\nonumber\\
&&4\pi\delta J^{\phi}=\frac{\alpha}{\Sigma}[\delta B_{\theta,r}+a_r\delta B_{\theta}]\nonumber\\
&&-\frac{1}{\alpha}[\delta E_{,t}^\phi+\omega\delta E_{,\phi}^\phi]-\frac{\delta E^r}{\alpha}(\Gamma_{t r}^\phi+\omega\Gamma_{\phi r}^\phi-\omega a_r-\alpha\gamma_{rr}\sigma^{\phi r}).\nonumber\\
\end{eqnarray}
Since we need the covariant components of Eqs(\ref{p5b}) we find
\begin{eqnarray}\label{p5c}
&&4\pi \delta J_r=-(\frac{\Sigma^2}{\alpha A})\delta B_{,\phi}^{\theta}-(\frac{\Sigma^2}{\alpha A})\sigma_{r\phi} B^{\theta}\delta\upsilon^r-D_{\tau} \delta E_r,\nonumber\\
&&4\pi \delta J_{\theta}=0,\nonumber\\
&&4\pi \delta J_{\phi}=(\frac{\Delta}{\alpha})[\delta B_{,r}^{\theta}+\frac{2 r}{\Sigma} \delta B^{\theta}]\nonumber\\
&&+(\frac{\Sigma}{\alpha}) d^r B^{\theta}-(\frac{\Delta}{\alpha})\sigma_{\phi r} B^{\theta}\delta \upsilon^{\phi}-D_{\tau}\delta E_{\phi}
\end{eqnarray}

Further, eqs.~(\ref{p8}) reduce to the system
\begin{eqnarray}\label{f1p}
&&-(\frac{\Delta}{r^2})(\delta p)_{,r}=\frac{(\rho+p)}{\alpha}(n+i m\omega)\delta \upsilon^r\nonumber\\
&&+\frac{(\rho+p)}{\alpha}[G_2(r)+\alpha(\frac{A}{\Sigma})\sigma^{r\phi}](\frac{i}{m}\chi)\nonumber\\
&&-a^r\alpha(1+c_s^2)[\frac{n-i m\omega}{n^2+m^2\omega^2}][(1-c_s^2)\rho_{,r}\nonumber\\
&&-\frac{3}{4\pi}(B_{\theta}B_{,r}^{\theta}+\frac{2}{r} B^2+a_r B^2 ]\delta \upsilon^r\nonumber\\
&&+\frac{\alpha }{r^2}[\delta J_{\phi} B_{\theta}+J_{\phi}\delta B_{\theta}]
\end{eqnarray}
and
\begin{eqnarray}\label{f2x}
&&-(\frac{r^2}{A})(\delta p)_{,\phi}=\frac{(\rho+p)}{\alpha}(n+i m\omega)(\frac{i}{i m}\chi)\nonumber\\
&&+\frac{(\rho+p)}{\alpha}[G_1(r)+\alpha (\frac{\Sigma}{\Delta})\sigma^{\phi r}]\delta \upsilon^r\nonumber\\
&&-(\frac{\alpha }{r^2})B_{\theta}\delta J_r
\end{eqnarray}
where
\begin{eqnarray}\label{j1}
&&4\pi B_{\theta}\delta J_r=-(\frac{\Sigma^2}{\alpha A})B_{\theta}\delta B_{,\phi}^{\theta}-(\frac{\Sigma^2}{\alpha A})\sigma_{r\phi} B^2\delta\upsilon^r-B_{\theta}D_{\tau} \delta E_r,\nonumber\\
&&4\pi[B_{\theta}\delta J_{\phi}+J_{\phi}\delta B_{\theta}]=(\frac{\Sigma \Delta}{\alpha\Sigma})[B_{\theta}\delta B_{,r}^{\theta}+B_{,r}^{\theta}\delta B_{\theta}\nonumber\\
&&+\frac{4}{r} B^{\theta} \delta B_{\theta}+2a_r B_{\theta}\delta B^{\theta}]+(\frac{\Sigma\Delta}{\alpha\Sigma})\sigma_{r\phi}[\frac{i\chi}{m}] B^2-B_{\theta} D_{\tau}\delta E_{\phi},\nonumber\\
\end{eqnarray}
and because of the $D_{\tau} M^{\beta}\equiv M^{\beta}~_{;\mu}
U^{\mu}-U^{\beta}a_{\mu} M^{\mu}$, which is the Fermi derivative,
\begin{eqnarray}\label{j2}
&&B_{\theta}D_{\tau}\delta E_r=(\frac{\Sigma}{\alpha^2})[\frac{i\chi}{m}](n+i m\omega)B^2+(\frac{\Sigma}{\alpha^2})G_3(r) B^2\delta\upsilon^r,\nonumber\\
&&B_{\theta}D_{\tau}\delta E_\phi=-(\frac{\Sigma}{\alpha^2})(n+i m\omega) B^2\delta\upsilon^r-(\frac{\Sigma}{\alpha^2})[\frac{i\chi}{m}]G_2(r) B^2,\nonumber\\
\end{eqnarray}
We have defined above
\begin{eqnarray}\label{j3}
&&G_1(r)\equiv\Gamma_{t r}^{\phi}+\omega\Gamma_{r\phi}^\phi-a_r\omega\nonumber\\
&&=(\frac{\omega}{2r})[\frac{3r(r-2M)^2+a^2(r-4M)}{(r-2M)\Delta}]\nonumber\\
&&\equiv(\frac{\omega}{2r})\tilde{G}_1(r),\nonumber\\
&&G_2(r)\equiv\Gamma_{t\phi}^r+\omega\Gamma_{\phi\phi}^r\nonumber\\
&&=-(\frac{\omega}{2r})[\frac{(3r^2+a^2)\Delta}{r^2}]\nonumber\\
&&\equiv-(\frac{\omega}{2r})\tilde{G}_2(r),~~\mbox{~and~}\nonumber\\
&&G_3(r)\equiv G_1(r)+a_r\omega\nonumber\\
&&=(\frac{\omega}{2r})[\frac{3r^2-4Mr+a^2}{\Delta}]\nonumber\\
&&\equiv(\frac{\omega}{2r})\tilde{G}_3(r).
\end{eqnarray}
From eqs.~(\ref{f1p}) and (\ref{f2x}), using eqs.~(\ref{f1})-(\ref{j3}), we find
a complex equation of the form
\begin{equation}\label{z5}
\Lambda_1+i\Lambda_2=(N_1)_{,r}+i (N_2)_{,r}
\end{equation}
where
\begin{eqnarray}\label{y4}
\Lambda_1&\equiv& n(\frac{1}{\alpha\Delta})(\rho+p+\frac{B^2}{4\pi})(r^2\delta \upsilon^r)  \nonumber\\
&&-(\frac{\alpha a^r}{\Delta})[\frac{n}{n^2+m^2\omega^2}][(1-c_s^4)\rho_{,r}\nonumber\\
&&-(\frac{3}{4\pi})(1+c_s^2)(B_{\theta}B_{,r}^{\theta}+\frac{2}{r}B^2 +a_r B^2)]
(r^2\delta \upsilon^r)\nonumber\\
&&(\frac{2}{4\pi})(\frac{1}{r}+a_r)(B^{\theta}\delta B_{\theta})
\end{eqnarray}
\begin{eqnarray}\label{y5}
\Lambda_2&\equiv&( m\omega)(\frac{1}{\alpha\Delta})(\rho+p+\frac{B^2}{4\pi})(r^2\delta \upsilon^r)\nonumber\\
&&+\frac{(\rho+p+\frac{B^2}{4\pi})}{\alpha \Delta}[G_2(r)+(\frac{\alpha A}{r^2})\sigma^{r\phi}](\frac{r^2}{ m}\chi)\nonumber\\
&&+(\frac{\alpha a^r}{\Delta})[\frac{ m\omega}{n^2+m^2\omega^2}[(1-c_s^4) \rho_{,r}\nonumber\\
&&-(\frac{3}{4\pi})(1+c_s^2)(B_{\theta}B_{,r}^{\theta}+\frac{2}{r}B^2+a_r B^2)](r^2\delta \upsilon^r)\nonumber\\
&&(\frac{2}{4\pi})(\frac{1}{r}+a_r)(B^{\theta}\delta B_{\theta})
\end{eqnarray}

\begin{eqnarray}\label{z4}
N_1&\equiv&(\frac{nr^2\chi}{m^2})(\frac{1}{\alpha})(\frac{A}{r^4})(\rho+p+\frac{B^2}{4\pi})\nonumber\\
&&+(\frac{1}{4\pi})B_{\theta}\delta B^{\theta}
\end{eqnarray}
and
\begin{eqnarray}\label{z4b}
N_2&\equiv&( m\omega)(\frac{1}{\alpha})(\frac{A}{r^4})(\frac{r^2\chi}{m^2})(\rho+p+\frac{B^2}{4\pi})\nonumber\\
&&-(\frac{1}{m})(\frac{A}{r^4})\frac{(\rho+p)}{\alpha}G_1(r)(r^2\delta\upsilon^r)\nonumber\\
&&-(\frac{1}{m})(\frac{A}{r^2\Delta})\sigma^{\phi r}(\rho+p+\frac{B^2}{4\pi})(r^2\delta \upsilon^r)\nonumber\\
&&-(\frac{1}{m})(\frac{1}{\alpha})(\frac{A}{r^4}G_3(r))(\frac{B^2}{4\pi})(r^2\delta\upsilon^r)
\end{eqnarray}
Obviously, from eq.~(\ref{z5}) we have the following two equations
\begin{equation}\label{f1}
\Lambda_1=(N_1)_{,r}
\end{equation}
and
\begin{equation}\label{f2}
\Lambda_2=(N_2)_{,r}
\end{equation}
In the main text we consdier only eq.~(\ref{f1}) since eq.~(\ref{f2}) does not give any new and
significantly different results. Using eqs.~(\ref{z4b}) and (\ref{y4}), eq.~(\ref{f2}) may 
be written as
\begin{eqnarray}\label{second}
&&-F(r)\{(\frac{A}{r^4})(\rho+p+\frac{B^2}{4\pi})(r^2\delta\upsilon^r)_{,r}\nonumber\\
&&-(\frac{A}{r^4})[\frac{\tilde{G}_1(r)}{2r}(\rho+p+\frac{B^2}{4\pi})+a_r\frac{B^2}{4\pi}](r^2\delta\upsilon^r)\nonumber\\
&&-\lambda_{\sigma}\frac{r^2(3r^2+a^2)}{2A}(\rho+p+\frac{B^2}{4\pi})\}\nonumber\\
&&+[(\frac{A}{r^4})(\rho+p+\frac{B^2}{4\pi})(r^2\delta\upsilon^r)_{,r})]_{,r}\nonumber\\
&&-\{(\frac{A}{r^4})[\frac{\tilde{G}_1(r)}{2r}(\rho+p+\frac{B^2}{4\pi})+a_r(\frac{B^2}{4\pi})](r^2\delta\upsilon^r)\nonumber\\
&&-\lambda_{\sigma}\frac{r^2(3r^2+a^2)}{2A}(\rho+p+\frac{B^2}{4\pi})\}_{,r}\nonumber\\
&&=\frac{m^2}{\Delta}(r^2\delta\upsilon^r)(\rho+p+\frac{B^2}{4\pi})-[\frac{\tilde{G}_2(r)}{2r\Delta}\nonumber\\
&&+\lambda_{\sigma}\frac{r^2(3r^2+a^2)}{2A}](\rho+p+\frac{B^2}{4\pi})(r^2\delta\upsilon^r)_{,r}\nonumber\\
&&+[\frac{m^2}{n^2+m^2\omega^2}](\frac{\alpha^2 a^r}{\Delta})[(1-c_s^4)\rho_{,r}-\frac{1}{4\pi}(1+3c_s^2)B_{\theta}B_{,r}^{\theta}\nonumber\\
&&-(\frac{3}{4\pi})(1+c_s^2)(\frac{2}{r}+a_r) B^2\nonumber\\
&&+(\frac{2\Delta}{r a^r})(\frac{1}{4\pi})(B_{\theta}\delta B^{\theta})_{,r}](r^2\delta\upsilon^r)\nonumber\\
\end{eqnarray}
where $\tilde{G}_1(r)$, $\tilde{G}_2(r)$  are given by eq.~(\ref{j3}) $\lambda_{\sigma}$ is a constant that is related to the shear $\sigma^{r\phi}$ term and $F(r)$ is
\begin{eqnarray}\label{l1}
F(r)\equiv\frac{1}{A\Delta}[3r^5-5M r^4+4a^2 r^3+a^4 r-4M^2 a^2 r+Ma^4]\nonumber\\
\end{eqnarray}

\end{document}